\documentclass[manuscript,screen]{acmart}
\AtBeginDocument{%
  \providecommand\BibTeX{{%
    \normalfont B\kern-0.5em{\scshape i\kern-0.25em b}\kern-0.8em\TeX}}}

\setcopyright{acmcopyright}
\copyrightyear{2018}
\acmYear{2018}
\acmDOI{XXXXXXX.XXXXXXX}

\acmConference[Conference acronym 'XX]{Make sure to enter the correct
  conference title from your rights confirmation emai}{June 03--05,
  2018}{Woodstock, NY}
\acmBooktitle{Woodstock '18: ACM Symposium on Neural Gaze Detection,
 June 03--05, 2018, Woodstock, NY} 
\acmPrice{15.00}
\acmISBN{978-1-4503-XXXX-X/18/06}

\usepackage{xcolor}
\usepackage{fontawesome}
\usepackage{soul}
\usepackage{multirow}
\usepackage{graphicx}
\usepackage{comment}
\usepackage{csquotes}

\begin{document}

\title [Teachers' Information Needs for ChatGPT]{More than Model Documentation: Uncovering Teachers' Bespoke Information Needs for Informed Classroom Integration of ChatGPT} 


\author{Mei Tan}
 \affiliation{%
  \institution{Stanford University}
   \country{USA}
 }
 \email{mxtan@stanford.edu}

 \author{Hariharan Subramonyam}
 \affiliation{%
  \institution{Stanford University}
   \country{USA}
 }
 \email{harihars@stanford.edu}

\renewcommand{\shortauthors}{tan, et al.}

\begin{abstract}
ChatGPT has entered classrooms, but not via the typical route of other educational technology, which includes comprehensive training, documentation, and vetting. Consequently, teachers are urgently tasked to assess its capabilities to determine potential effects on student learning and instruct their use of ChatGPT. However, it is unclear what support teachers have and need and whether existing documentation, such as model cards, provides adequate direction for educators in this new paradigm. By interviewing 22 middle- and high-school teachers, we connect the discourse on AI transparency and documentation with educational technology integration, highlighting the critical information needs of teachers. Our findings reveal that teachers confront significant information gaps, lacking clarity on exploring ChatGPT's capabilities for bespoke learning tasks and ensuring its fit with the needs of diverse learners. As a solution, we propose a framework for interactive model documentation that empowers teachers to navigate the interplay between pedagogical and technical knowledge. 
\end{abstract}

\maketitle

\section{Introduction}
ChatGPT~\cite{chatgpt}, a conversational artificial intelligence interface using the latest advancements in natural language processing, has presented \textit{immediate} and \textit{significant} impacts on the education domain. Initial concerns with ChatGPT centered around issues of academic integrity and plagiarism and their threat to conventional forms of educational assessment~\cite{stokel2022ai}. Consequently, school districts responded by blocking access to the website from their schools' networks~\cite{rosenblatt2023chatgpt}. However, the ineffectiveness of such bans and their subsequent repeals \cite{singer2023} pushed the narrative among educators to evolve and consider opportunities to positively apply it to teaching and learning. Across teacher-facing websites, there is a surge of ad-hoc high-level suggestions on leveraging ChatGPT to assist teachers by generating lesson plans, reading passages, assessment questions, models of student work, etc \cite{chatgptteachers}. At the same time, teachers are advised of its risks and limitations, cautioning them to adopt a thoughtful, well-researched, and purposeful approach to integrating ChatGPT in educational settings—i.e., \textit{informed classroom integration}.

Educators are \textit{tasked} to test ChatGPT's capabilities to determine potential effects on student learning~\cite{mishra2019considering}, devote time to oversight and prevention of misuse, and devise a clear strategy for facilitating student use of the tool~\cite{mittechreview}. Such an evaluation necessitates that teachers develop competencies to understand the technology, recognize its vulnerabilities, and adeptly leverage its capabilities in alignment with pedagogical goals~\cite{kasneci2023chatgpt}. However, this is easier said than done. Unfortunately, educators and educational institutions often lack the knowledge or expertise to effectively integrate new technologies in their teaching~\cite{hew2007integrating, baylor2002factors}. Teachers lack training in technology and technical support~\cite{francom2020barriers, hsu2016examining}. The act of integrating technology is time-consuming and interrupts instruction, requiring additional planning and new routines to implement technology-integrated lessons~\cite{KOPCHA20121109}. 

Compared with the integration of conventional educational technology, that of ChatGPT introduces more technical complexities and fewer facilitating resources. To integrate conventional tools, educators and school administrators often work with the product or service companies to assess infrastructural needs, compatibility with existing systems, and cost~\cite{roblyer2007integrating}. They may even engage in a pilot testing phase to ensure its alignment with educational objectives while emphasizing ethical considerations, particularly concerning student data and privacy~\cite{ertmer2010teacher}. During this process, companies provide formal training, documentation, continued technology support, and established processes for troubleshooting failures. Unfortunately, this is not the case with ChatGPT. Further, while most educational software is designed to perform well-defined tasks in specified use contexts, ChatGPT is general-purpose; it can ``write stories, give life advice, even compose poems and code computer programs~\cite{cooper2023examining}.'' Emergent behaviors of language models present new challenges for fairness, accountability, and transparency \cite{weidinger2021ethical}. Documentation from OpenAI\footnote{Since the release of ChatGPT in November 2022, available documentation consisted of the GPT-3 model card \cite{gpt3modelcard} and a since-deprecated page of notes for educators that introduced potential opportunities for applications in education and documented similar disclosures to those in the model card. In August 2023 OpenAI replaced this page with the release of its Teaching with AI guide \cite{openai_teachingguide} and FAQ \cite{openai_faq}.} has disclosed the risks of ChatGPT, including its propensity to produce false information and content that perpetuates harmful biases and stereotypes~\cite{gpt3modelcard, openai_faq}. However, established frameworks for AI transparency \cite{mitchell2019model} cater predominantly to technical experts, emphasizing the technology's capabilities and overlooking the practical challenges and contexts faced by practitioners \cite{boyarskaya2020overcoming, konigstorfer2021software}. As a result, documentation may fail to support teachers' abilities to grasp ChatGPT's underlying intricacies, discern valuable uses for its features, and navigate its potential pitfalls.

In this work, we investigate teachers' information needs regarding the integration of ChatGPT. We interview experienced teachers to understand their evaluation of available information, their approach to information seeking, and their strategies for capability testing in the ChatGPT interface. In doing so, we identify critical information gaps limiting effective integration. Our findings reveal that teachers face challenges discovering applications of ChatGPT relevant to their pedagogical goals, struggling to contextualize its technical affordances and limitations in classroom practice. Teachers lack competencies in prompt engineering and approach experimentation with faulty mental models and narrowly defined goals, resulting in flawed outputs that limit perceptions of ChatGPT's utility in education. We highlight the gaps in existing documentation practices that inadequately address these challenges and present arguments for experimentation-based documentation tools that support dynamic and interactive understanding. Our key contributions include taxonomies of teachers' information needs and experimentation behaviors, as well as a proposed framework to support teachers in exploring---rather than reading about---the technical capabilities of ChatGPT.

\section{Related Work}

\subsection{AI in Education}
The increasing presence of artificial intelligence in educational contexts presents novel pedagogical opportunities and disruptions~\cite{hwang2020vision, celik2022promises, chen2022two}. AI capabilities in 
learning analytics, eye-tracking, speech recognition, computer vision, and natural language interaction have the potential to enrich student learning and complement the work of teachers~\cite{reiss2021use, niemi2023ai, roschelle2020ai}. Adaptive learning systems~\cite{colchester2017survey}, text-based chatbots~\cite{chocarro2023teachers}, and intelligent tutoring systems~\cite{ma2014intelligent} provide students with personalized and optimized learning paths, catering to specific student requirements, habits, and abilities~\cite{bhutoria2022personalized, hwang2014definition}. Other AI-based tools facilitate teachers in monitoring student progress~\cite{keuning2021differentiated} and providing just-in-time feedback and assessment~\cite{luckin2016intelligence}. Large language models, including ChatGPT~\cite{team2022chatgpt}, have demonstrated use across a wide range of language tasks, from translation and question answering to writing essays and computer programs~\cite{kasneci2023chatgpt}. Researchers have applied large language models to specific pedagogical tasks, for example, as a generator of assessment questions~\cite{dijkstra2022reading}, code explanations~\cite{macneil2022generating}, and feedback~\cite{jia2021all}, and as a conversational partner for foreign language practice~\cite{ji2023systematic}. 

At the same time, AI technology has introduced new uncertainties to the adoption and integration process~\cite{luckin2022ai}. As AI-based technologies increasingly perform academic and instructional tasks~\cite{tack2022ai}, they recalibrate classroom roles and expectations. Teachers fear displacement and relegation to the role of monitoring social and emotional needs~\cite{pinkwart2016another, hao2019china, guilherme2019ai}. Though other works have argued for the irreplaceable role of the teacher~\cite{cheng2019case}, designing tools and classroom practices for complementarity between teachers and AI remains a persistent challenge~\cite{holstein2019designing, schiff2021out}. 

Specific to ChatGPT, it remains unclear how teachers navigate and make pedagogical advantage of such open-ended generative systems~\cite{zawacki2019systematic, luckin2022ai}. In response, researchers have called for teachers to develop new literacies to understand the technology driving AI and large language models~\cite{kasneci2023chatgpt} and pedagogical strategies for integration~\cite{zawacki2019systematic}. Teachers and administrators also need competencies in evaluating AI in terms of its limitations and ethical issues~\cite{kim2021analyzing, celik2023towards}, which include biased outputs~\cite{baker2021algorithmic} and issues of fairness~\cite{almusharraf2023error}, data privacy concerns~\cite{potgieter2020privacy}, and unequal access~\cite{reich2017good}. Such technical knowledge would enable practitioners to interpret documents explaining the behaviors of systems, discover and prevent bias, and demand accountability for fairness and transparency~\cite{bogina2021educating}.

\subsection{Educational Technology Adoption}
Building on the concerns mentioned above, effective integration of educational technology remains a perennial challenge for teachers~\cite{hixon2009revisiting, cuban1986teachers}. An extensive body of research is dedicated to identifying the factors that limit its classroom adoption and utilization~\cite{levin2008teachers}. These factors include teachers' age and experience~\citep{inan2010factors}, time constraints~\cite{wepner2003three}, pedagogical beliefs~\cite{tondeur2017understanding}, limited technology-related knowledge and skills~\cite{seufert2021technology}, and insufficient training~\cite{tondeur2012preparing}.  

General technology acceptance frameworks have commonly been applied in the education context~\cite{granic2022educational} to model teachers' technology adoption practices. The Technology Acceptance Model (TAM) explains users' behavioral intention toward technology as a product of its perceived usefulness and perceived ease of use~\cite{davis1987user}. The perceived ease of use of technology is influenced by its quality, complexity, and accessibility~\cite{teo2009modelling}. This perception, in turn, significantly affects its perceived usefulness~\cite{davis1987user, joo2018factors}. Extended models, such as the Unified Theory of the Acceptance and Use of Technology (UTAUT)~\cite{bervell2017validation}, have demonstrated the role of teachers' social and physical environments. Studies applying these models have observed the impact of district resource factors, teachers' technology anxiety~\cite{henderson2021teacher}, and perceptions of subjective norms among peers~\cite{cheung2013predicting}. 

Beyond the initial uptake of technologies, several frameworks developed in the education context aim to describe the application of these tools in practice. The notion of technology integration is multi-dimensional, loosely defined, and operationalized in research through numerous instruments~\cite{consoli2023technology}. The SAMR model evaluates teachers' effective technology integration by characterizing the degree to which technologies replace or amplify traditional practice~\cite{hamilton2016substitution}. The PICRAT model additionally captures students' passive and interactive engagement with classroom technologies ~\cite{kimmons2020picrat}. Other models emphasize the pedagogical alignment of technology use in teaching methods and curriculum~\cite{welsh2011florida, koehler2009technological}. 

Perhaps the most prominent model of technology integration in education is the TPACK, which models the complex integration of technological (T), pedagogical (P), and content knowledge (CK)~\cite{koehler2009technological}. Extensions of TPACK additionally address the role of contextual knowledge for adapting technology use to individual students and classrooms~\cite{mishra2019considering}. The framework provides an explicit mechanism for discussing technology integration in service of teaching and learning~\cite{baran2011tpack}.  Studies applying the TPACK have found that teachers struggle to develop sufficient pedagogical content knowledge prior to technology integration~\cite{pamuk2012understanding}, leading to inadequate consideration of curricular issues. Teachers' rationales for technology adoption are often disconnected from improving content understanding~\cite{graham2012using}. This discrepancy constrains teachers' design of technology-integrated activities and perpetuates barriers to integration~\cite{karchmer2023mixed}. New waves of technological developments, including the emergence of artificial intelligence, place additional demands on teachers' technology-related knowledge, skills, and attitudes~\cite{seufert2021technology, aldunate2013teacher}.  Recent applications of TPACK have noted the importance of technological knowledge in enabling teachers to interact with AI-based tools to assess and better understand their pedagogical contributions~\cite{celik2023towards}. 

\subsection{Technology Documentation}
Despite calls for educating teachers on fairness, accountability, transparency, and ethics in AI~\cite{bogina2021educating, holmes2021ethics, ng2021conceptualizing}, teachers are provided limited information and continue to face difficulties understanding AI behaviors~\cite{castelvecchi2016can}. Though documentation is a central mechanism for ensuring the fair and accountable application of AI in practice, recent research has shown that current documentation guidelines do not provide sufficient guidance for practitioners~\cite{konigstorfer2021software}. Insufficient documentation is an essential barrier to the adoption of AI~\cite{konigstorfer2022ai}. While some non-adopters are cautious of the risks of limited explainability, others perceive documentation as a burden to consume.

Researchers have argued that documentation for software and AI are subject to different requirements~\cite{konigstorfer2021software}. Software documentation describes what a software system does, how it operates, and how it should be used~\cite{aghajani2019software}. Documents include technical reference guides for developers and auditors and user manuals for practitioners. Though such documentation is essential to validate the proper functioning of software tools, studies have found problems negatively impacting their quality and usefulness, including incomplete or out-of-date information, missing definitions, low readability, and technical jargon~\cite{aghajani2020software, zhi2015cost, 8811931}.

A smaller body of work is dedicated to the development of AI documentation and reporting frameworks to make AI systems more transparent and help users establish trust. Researchers have proposed factsheets~\cite{arnold2019factsheets}, datasheets~\cite{gebru2021datasheets, holland2020dataset, bender2019data}, and model cards~\cite{mitchell2019model} to specify the content of AI documentation, including purpose, performance metrics, training data composition, and disclosures of limitations. Though model cards were originally proposed to accommodate a spectrum of stakeholders~\cite{mitchell2019model}, their highly technical text-based format leaves many without AI expertise underserved~\cite{crisan2022interactive}. Studies have emphasized the importance of practitioner interaction with AI models and their data~\cite{amershi2014power, crisan2022interactive}, and companies such as HuggingFace and Google Cloud have implemented interactive modalities for assessing model performance. 

However, there is a lack of research evaluating the degree to which existing AI documentation is understandable and applicable for end-users~\cite{piorkowski2020towards}. To support the effective integration of AI-based tools in education, researchers must investigate the delta between available documentation frameworks and teachers' contextual information needs. To this end, we ask the following research questions: 
\begin{itemize}
    \item RQ1: How do teachers seek and evaluate available information about ChatGPT to motivate integration decisions?
    \item RQ2: How do teachers assess the technical capabilities of ChatGPT for pedagogical utility?
    \item RQ3: How do existing AI transparency and documentation frameworks fall short in addressing the information needs of teachers for the effective integration of ChatGPT into classroom settings?
\end{itemize}
\section{Method}
To investigate how teachers approach information gathering and assessment of ChatGPT for classroom integration, we conducted semi-structured interviews with experienced teachers. Each interview consisted of discussions around participants' experiences with general technology adoption and the appearance of ChatGPT in education contexts, an evaluation of available transparency documentation regarding the technology behind ChatGPT, and experimentation with the ChatGPT interface. Interviews were held virtually via Zoom and lasted between 90 and 120 minutes. Participation was voluntary, and all participants were compensated with \$100 for their involvement. Our institution’s IRB approved the study. 

\subsection{Participants}
We recruited experienced teachers through our university alumni networks and also reached out to school districts engaged in long-standing partnerships with our university. Due to the central functionality of text generation in the ChatGPT interface, we recruited middle- and high-school teachers for whom student writing is central to their instructional practice. We selected 22 participants, meeting our selection criteria by order of sign-up. The participants include 9 teachers with experience teaching at the middle school level and 19 teachers with experience teaching at both middle and high school levels. Ten teachers have between two and five years of experience, nine teachers have between six and ten years of experience, and three teachers have more than ten years of experience. Their teaching experiences involve a variety of writing-intensive subjects, including English Language Arts, Social Studies, History, and Humanities. 

\begin{table}[]
\resizebox{\columnwidth}{!}{%
\begin{tabular}{ccccc}
\toprule
\textbf{Phase} & \textbf{Teacher ID} & \textbf{Years of Experience} & \textbf{Grade Levels} & \textbf{Subjects}     \\ \midrule
\multirow{14}{*}{\rotatebox[origin=c]{90}{Phase 1: Open}} & 1                   & 4                            & 10-11                 & English Language Arts \\
 & 2  & 2  & 9       & English Language and Composition                                    \\
 & 3  & 2  & 7       & English Language Arts, Social Studies                               \\
 & 4  & 3  & 9-12    & Social Studies                                                      \\
 & 5  & 3  & 7-8, 12 & English Language Arts                                               \\
 & 6  & 19 & 7-12    & English Language Arts                                               \\
 & 7  & 9  & 7-12    & English Language Arts                                               \\
 & 8  & 6  & 9-12    & English Language Arts, Humanities                                   \\
 & 9  & 7  & 9-12    & English Language Arts, AVID                                         \\
 & 10 & 13 & 7-12    & AP Government, World History                                        \\
 & 11 & 7  & 7-12    & English Language Arts                                               \\
 & 14 & 9  & 7-8     & English Language Arts                                               \\
 & 15 & 4  & 7-8     & English Language Arts                                               \\
 & 16 & 9  & 9-12    & Social Studies, Human Geography, U.S. History                       \\ \midrule
\multirow{8}{*}{\rotatebox[origin=c]{90}{Phase 2: Guided}}  & 12                  & 3                            & 9-11                  & English Language Arts \\
 & 13 & 5  & 9-12    & AP Language and Composition                                         \\
 & 17 & 12 & 9-12    & English Language Arts, Humanities, Social Studies, AP U.S. History  \\
 & 18 & 2  & 9-12    & Humanities                                                          \\
 & 19 & 5  & 9-12    & English Language Development                                        \\
 & 20 & 6  & 10-11   & English Language Arts                                               \\
 & 21 & 8  & 7-12    & English Language Arts, ESL, Reading Intervention, Special Education \\
 & 22 & 6  & 9-12    & English Language Development                                        \\ \bottomrule
\end{tabular}%
}
\caption{Each interview is listed by teacher ID with years of experience, grade levels, and subjects taught. Phase indicates iteration of the experimentation portion of the protocol, where some interviews engaged teachers in open exploration of ChatGPT and its potential classroom applications while others guided teachers to assess ChatGPT in the context of a specific use case.}
\label{tab:my-table}
\end{table}

\subsection{Protocol}
In line with the iterative nature of qualitative research, our interview protocol underwent periodic revisions as the study progressed~\cite{charmaz2014constructing, creswell2017research}. This was essential to ensure that the questions being posed were not only reflective of our evolving understanding of the research context but also responsive to the emerging needs and insights from the participants. Initial interviews were guided by a preliminary set of questions developed from literature and involved open-ended exploration (phrase 1, with 14 participants). However, as patterns began to emerge and new themes were identified, the protocol was adjusted to delve deeper into specific areas of interest and to better support participant engagement (phase 2, with 8 participants). All participants engaged in sharing their initial experiences with ChatGPT and evaluating model card information. Here we discuss each of these steps in detail. 

\subsubsection{Experience Sharing}
The semi-structured interviews started with a discussion of participants’ experiences with classroom technology integration. The interviewer first asked participants to share contextual factors, such as hardware distribution and access, administrative oversight and decision-making, information channels, and professional development opportunities. Then, teachers were asked to reflect on the information they seek and the criteria they use to assess whether a technology application is supportive of student learning and how they would approach incorporating the application into their instructional practice. The interview questions then shifted to focus on AI applications and ChatGPT. The interviewer asked teachers to discuss student, teacher, and administrative responses to the emergence of ChatGPT and probed about perceived opportunities and difficulties. Teachers answered questions about the information and narratives they have consumed regarding ChatGPT and their degree of personal interaction with the tool.

\subsubsection{Model Card Evaluation}
The interviewer introduced the model card for GPT-3 \cite{gpt3modelcard} as a technical document for the technology driving ChatGPT and explained that model cards~\cite{mitchell2019model} address transparency needs for machine learning models by documenting training data composition, performance accuracy, and limitations. Note that education-specific documentation from OpenAI was not yet available at the time of interviews (e.g., the Teaching with AI guide~\cite{openai_teachingguide} and Educator FAQ~\cite{openai_faq}); however, these resources adapt a subset of technical disclosures described in model documentation, and we discuss their limitations in Section~\ref{sec: discussion}. Teachers were primed to expect the ``technical language'' and given time to read the document. They were subsequently asked to reflect on the information presented. The interviewer prompted teachers to identify the most meaningful disclosures from the model card for the education domain and explain their reasoning for how the information would affect their classroom practice. Teachers were then asked to identify the disclosures that are not applicable, understandable, or actionable for the education domain and discuss outstanding information needs not addressed by the model card.

\subsubsection{Phase 1: Open Exploration}
In the first 14 interviews, teachers were asked to engage in open exploration and information-seeking regarding ChatGPT. The interviewer described the motivation for the task: district administrators are seeking teacher perspectives to inform the design of policies overseeing the use of ChatGPT. Teachers were asked to determine whether and how they would integrate the use of ChatGPT in their classroom practice in preparation for responding to the district request with a well-reasoned report. They were asked to share their browser screens and narrate their thought processes as they searched for necessary information and interacted with the ChatGPT interface. During the process, the interviewer encouraged teachers to lead the exploration and only engaged the teachers in discussion about the quality of ChatGPT responses and answered questions to facilitate teachers' interactions. After observing the challenges teachers faced while independently directing exploration with ChatGPT, in the final eight interviews, we followed an adapted protocol that omitted the open experimentation task and instead introduced a guided experimentation task (Phase 2 sessions).

\begin{figure}[h]
\includegraphics[width=\textwidth]{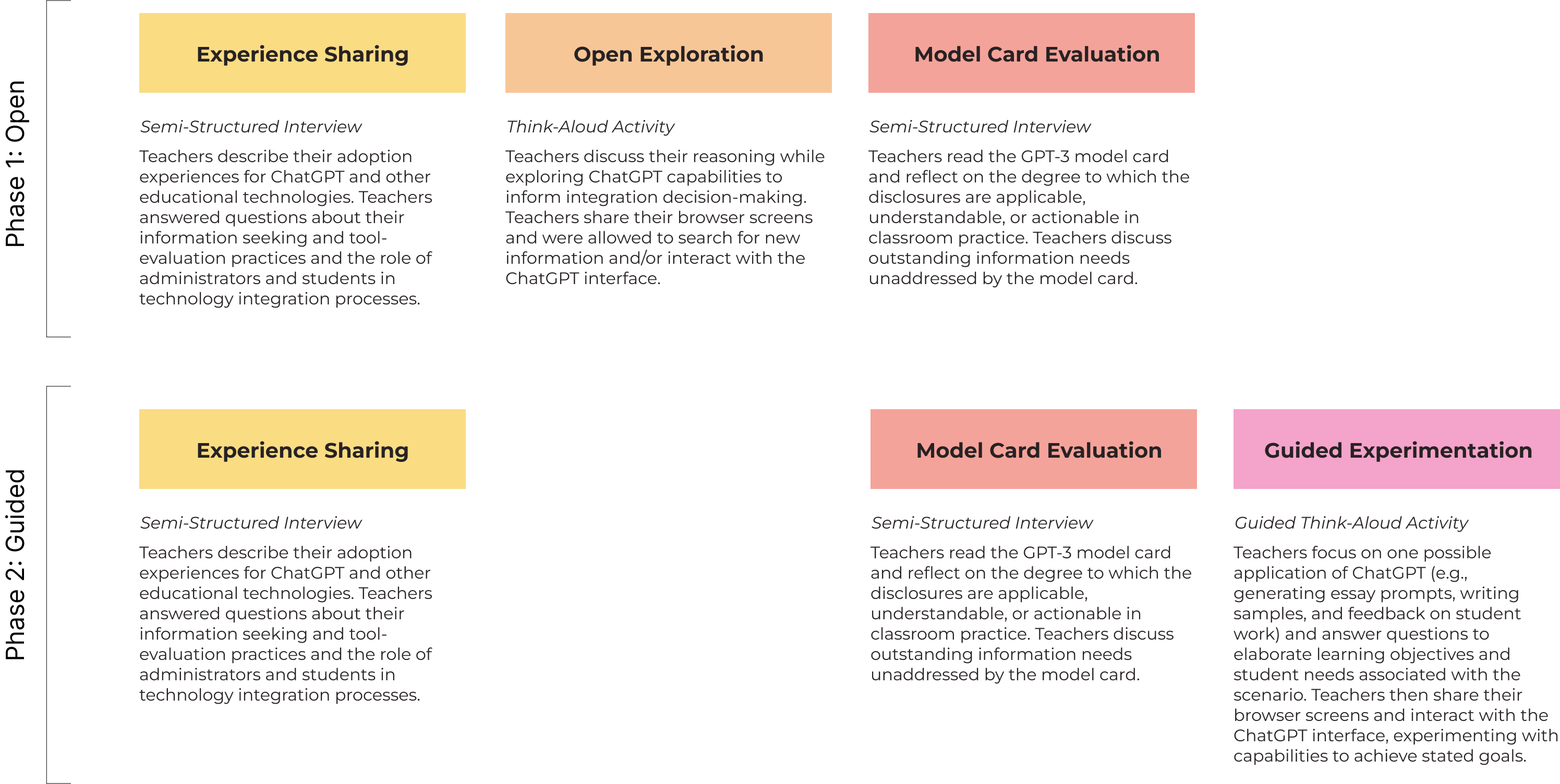}
\caption{Components of our interview protocols across two phrases. Both phases engaged teachers in experience sharing and model card evaluation, but the second phase replaced open exploration with a guided experimentation think-aloud activity.}
\label{fig:framework}
\end{figure}

\subsubsection{Phase 2: Guided Experimentation}
For the final eight interviews, a guided experimentation task followed model card evaluation. Teachers were presented with a list of possible teacher-facing and student-facing applications of ChatGPT that teachers had suggested in previous interviews (e.g., generating essay prompts, writing samples, and feedback on student work). After instructing teachers to select one application as a focal scenario for the experimentation activity, the interviewer led teachers through a series of questions eliciting the student needs and learning goals associated with the scenario. Teachers were asked to describe in detail their approach to the scenario without ChatGPT, including differentiated learning needs, pedagogical considerations, evaluation criteria, and associated classroom management needs. Next, the interviewer instructed teachers to consider how ChatGPT could be used to support the learning scenario. They were asked to share their browser screens and narrate their thought process as they interacted with the ChatGPT interface, using the needs elicited through the learning context discussion to guide experimentation. During the experimentation process, the interviewer guided teachers to test ChatGPT capabilities related to their stated learning needs. 

\subsubsection{Debrief}
The interviews concluded with a series of debrief questions related to the perceived overuse, misuse, and harms of integrating ChatGPT in the classroom, based on what teachers had gathered from documentation and experimentation. The interviewer prompted teachers to articulate the remaining gaps in information and support required to minimize the risk of harm.

\subsection{Analysis}
The interview transcripts include question-answering and think-aloud data. We first manually verified the text of the automated transcriptions from Zoom against the video recordings, then conducted inductive qualitative coding in Atlas.ti~\cite{atlas.ti} using a grounded theory approach~\cite{strauss1990basics} beginning with in-vivo analysis. During coding, we also had the video synchronized with the transcript and added screenshots of prompts from the video into the transcript. The two authors independently open-coded a common set of three interviews and collaboratively developed an initial codebook, which the authors applied to analyze the remaining transcripts. The coding scheme included references to information seeking (sources, framing, attitudes, mental models, gaps, etc.), experimentation (prompting strategies, response validation, etc.), and integration decision-making (support needs, trade-offs, output characteristics, etc.). Throughout the coding process, the authors wrote reflective memos describing emerging themes and making connections across interviews. The authors then discussed codes and memos, grouping codes to identify higher-level themes and synthesizing findings. 
\section{Findings}
In each interview, we engaged teachers in discussions and activities to assess their information-seeking and experimentation approaches to ChatGPT. In doing so, we identify information gaps, experimentation goals and strategies, and barriers to integration. We summarize our study findings in terms of (1) information seeking and comprehension, (2) capability exploration and experimentation, and (3) classroom integration support needs. 

\subsection{Information Seeking and Comprehension}
We observed teachers' information-seeking processes and examined how teachers make sense of available information regarding ChatGPT. Through our discussions about their prior experiences, we discovered that the adoption process for ChatGPT significantly diverged from typical educational technologies, marked by a scarcity of practical guidance and unconventional avenues of information access. Teachers reported disengagement when sifting through web articles and static documentation, citing frustrations due to time constraints, and the irrelevance of the available information to classroom settings. Here we describe teachers' information sources and document their contribution to information gaps and flawed mental models during comprehension. 

\subsubsection{Information Sources}
Nearly all teachers considered ChatGPT their first exposure to AI tools in education. Participants recalled initial discussions prompted by major news outlets at the time of release. Information shared among colleagues in the school setting was dominated by discussion of news articles and perspectives from the media. While most early conversations occurred organically among friends, in a few cases, teachers organized ad-hoc department meetings. Teacher T5 recounted that these sessions remained primarily focused on the media lens. Without coordinated administrative guidance or official professional development, teachers were encouraged to seek their own resources as needed. For many teachers, this need developed from increasing encounters with suspicious student work, later confirmed by students to be generated by ChatGPT. Teachers’ information-seeking led them to Twitter posts, podcasts, news sites, education blogs, and a variety of online articles. Some teachers described social media as a primary source of information. Teacher T2 explained:

\begin{displayquote}
T2: \textit{I'm on teacher TikTok and tech TikTok\ldots, so then the convergence of those is like, “Here's AI that teachers are using to make lesson plans or create images to go on their slides”. I also got the TikTok videos that are for students that are “Students, if you want to cheat on your essay use this”.}
\end{displayquote}

However, other teachers reported feeling disengaged from doing personal research regarding ChatGPT, citing time constraints. Indeed, information-seeking practices for ChatGPT are effortful and stand in stark contrast to teachers’ information access for other classroom technologies. Though typical technology adoption procedures vary, all teachers reported more district- and company-led support for integrating other tools. Teachers access user guides highlighted on educational technology company websites and videos of teacher walk-throughs demonstrating the interface. Some classroom technologies are accompanied by organized professional development sessions and administrative guidance on requirements, procedures, and learning goals. Teacher T1 described professional development sessions that are often facilitated by a representative from the educational technology company:

\begin{displayquote}
T1: \textit{We also have the option to meet with someone from Writable to like help us roll it out \ldots. It’s kind of like two main areas that we talk about. The first is just like the tech support: “How are students being presented this information? How is it integrating with Schoology, how do we need to be setting up our grade books to be compatible with it?” Troubleshooting and problem solving for like, “Okay this doesn’t work the way we want so how do we have to adjust?” And then the other aspect, of course, is like, “What do we want students to be doing in this platform?”}
\end{displayquote}

The availability of teacher-facing resources for conventional educational technologies goes beyond informing teachers about the tool and provides the necessary support and context-specific guidance to integrate it into classroom practice. Despite its presence in classrooms, ChatGPT is not a dedicated educational technology tool, and teachers lack formal training in its use. As a result, teachers fall back on unconventional sources and abstract information disconnected from their domain.

\subsubsection{Perceptions and Negotiated Understandings}
Based on available information, teachers arrived at a conflicted perception of ChatGPT. All teachers in our interviews described feeling overwhelmed by its harmful implications for academic integrity. Many discussed their experiences with detection tools, student confrontations, unsuccessful bans, and the loss of traditional methods of assessment. These negative perceptions are compounded by the volume of unknowns:

\begin{displayquote}
T21: \textit{It seems like education as a whole is very hesitant about it, and it's just like people are just going to use it to cheat. Shut it down. \dots And I really don't want education to change to just all we do is monitor to make sure kids aren't cheating, because it's not what education is about. \dots I think that it’s just unknown, like I don't know much about it. I don't know how it could be useful. I mean, there's some ways I could see how it could be detrimental in certain instances, but I think it's just like unknown.}
\end{displayquote}

Despite their reservations, teachers felt pressure to embrace ChatGPT to advance their teaching practice for fear of being left behind. These fears extend to ensuring that students have access to tools that may be advantageous for their future opportunities:

\begin{displayquote}
T3: \textit{Especially working in low-income areas where students like aren't having access to the tools that a lot of like their peers in different contexts like across the country or the world do \dots No, like we should have access to it too, so like our students can learn to use that too.}
\end{displayquote}

However, with limited information, teachers developed faulty mental models for understanding the capabilities of ChatGPT. In some sessions, teachers made assumptions about limitations in the style of text produced by ChatGPT and expressed confidence in their ability to differentiate the tone of students’ writing from that of ChatGPT. Perhaps drawing on the same misunderstanding that ChatGPT has a defined and identifiable way of writing, teachers expressed overconfidence in the accuracy of ChatGPT plagiarism detectors. Additionally, several teachers likened the advent of ChatGPT to that of Sparknotes and Wikipedia, framing ChatGPT's capabilities around information retrieval. Teacher T17 attributed the information in a ChatGPT-generated study guide for the AP U.S. History exam to a search of online materials published by Barron’s and the Princeton Review. The assumption that ChatGPT could perform Google searches and access websites, academic journals, and textbooks led teachers to express overconfidence in the validity of outputs. 

\begin{table}[]
\resizebox{\columnwidth}{!}{%
\begin{tabular}{p{4cm} p{9cm} p{9cm}}
\toprule
\textbf{Needs Category} & \textbf{Information Needs}                                                                                                                                                                                                                                             & \textbf{Deficits in Current Documentation}                                                                                                                                                                                                                                                 \\
\midrule
Data Policies           & Safety of student data inputs, privacy and security practices, incorporation into continued training, implications for student ownership of their work, FERPA compliance                                                                                               & Though OpenAI documents their privacy policies \cite{openai_privacy}, disclosures are undiscoverable, and teachers gather false information about data collection by directly asking ChatGPT.                                                                            \\
Content Restrictions    & Age restrictions, censorship of unsafe content, restriction of unsafe searches                                                                                                                                                                                         & Though the OpenAI privacy policy \cite{openai_privacy} and FAQ \cite{openai_faq} state that use by minors is against the terms of use, there are no documented age restrictions. Documentation does not specify content censorship practices.          \\
Oversight               & Teacher, parent, and administrative access to conversations, alerting of concerns for student safety surfaced through chats                                                                                                                                            & Documentation does not specify processes for oversight. The OpenAI privacy policy indicates emailing their legal team \textbackslash{}cite\{openai\_privacy\} in cases where minors share information with the system.                                                                     \\
Controling Misuse       & Restricting student misuse, unsafe interactions, guidelines for classroom management                                                                                                                                                                                   & Documentation does not specify guidelines for managing student misuse.                                                                                                                                                                                                                     \\
Academic Integrity      & Ethical standards for when and how it is appropriate for students to use ChatGPT, how to cite ChatGPT output and discover underlying sources, accuracy of tools to detect ChatGPT content                                                                              & Documentation discloses that inaccuracy of AI detectors \cite{openai_faq} and recommends having students share a record of their interactions with ChatGPT, but provides no further guidance about managing use in support of changing standards for academic integrity. \\
Educational Alignment   & Goals of integration, pedagogical relevance of ChatGPT, more exhaustive illustration of use cases and generative capabilities                                                                                                                                          & Though the OpenAI Teaching with AI guide cite{} and various online blogs and articles provide examples, they are few and abstract. Documentation does not discuss the breadth of pedagogical applications or the purposes of integrating ChatGPT in education.              \\
Information Access      & Relevance of training data to educational materials, access to textbooks and literary works, availability of historical and current information, internet search capabilities, filtering and preprocessing of training data to mitigate biases & Though the GPT-3 model card discloses training data sources, documentation does not contextualize data disclosures in terms of access to specific content or relevant learning materials. Documentation does not specify data pre-processing or filters upon the training data.            \\
Restrictions            & Scope of introduced filters, changing restrictions, filter-restricted content and actions                                                                                                                                                                              & Documentation does not specify modifications or scope of filter-based restrictions upon model or interface behavior.                                                                                                                                                                       \\
Performance             & Accuracy of classroom-relevant tasks such as summarizing literary or historical text, evaluation of student engagement and interface usability in educational contexts, disclosures of gaps in testing                                                                 & Documentation does not provide domain-relevant performance metrics.                                                                                                                                                                                                                        \\
Limitations Workarounds & How to assess the risk of encountering biased or inaccurate information, prompts and content to avoid in order to limit the risk of hallucination                                                                                                                              & Documentation does not provide guidance for avoiding and working around limitations.                                                                                                                                                                                                       \\
Prompting               & Prompt engineering practices, chat memory context, formatting inputs such as rubrics and passages of text                                                                                                                                                              & Documentation does not provide guidance for navigating the interface or effectively prompting ChatGPT.    \\
\bottomrule
\end{tabular}
}
\caption{Information needs expressed by participants in interviews and the degree to which they are addressed in current documentation.} 
\end{table}

\subsubsection{Information Gaps}
Across all interviews, teachers expressed frustration with the applicability of available information in the classroom. In addition to conceptual uncertainties about the underlying technology and its limitations, teachers asked questions about its use and regulation in academic and administrative settings. In Table ~\ref{infoneeds}, we present a taxonomy of information needs elicited by teachers in our interviews. Of these, teachers voiced the greatest information concerns about learning-relevant use cases and pedagogical goals of integration:

\begin{displayquote}
T6: \textit{I would be curious what goals the district would have for the use of the technology. Like, what would they want us to do with it? How they would recommend that we use it? More than just samples like, what are the applications that I'm kind of missing?}
\end{displayquote}

Teachers additionally voiced the need for support in managing classroom use, negotiating issues of academic integrity, adapting academic standards, preventing student misuse, and restricting student access to harmful content. In addition to general questions about the privacy and security of student data, teachers expressed concerns about administrative access to conversations to ensure student safety. Teacher T9 explained that these procedures are defined in other student writing contexts:

\begin{displayquote}
T9: \textit{Would there be any sort of like way for the school to know if kids were asking things that were worrisome, whether it's self-harm, weaponry, abuse, etc.? It'd probably be built-in in some way, but what is that way? I know that with like the AP Exams, if you're grading one and you come across something concerning there's a whole process to follow for the school to be alerted so like, what is it here?}
\end{displayquote}

When exposed to the GPT-3 model card, teachers expressed interest in its disclosures but struggled to contextualize the information in their instructional practices. Teachers noted the irrelevance of model performance metrics but inquired about ChatGPT's accuracy on classroom tasks, such as summarizing a reading comprehension article. Further, though the model card makes transparent that the dataset is composed of text posted to the internet, the relevance of training data to educational materials remains unclear. Teachers wondered whether ChatGPT has knowledge of specific textbooks, niche works of literature, academic journal papers, historical documents, and other texts commonly used in their practice. For some teachers, knowing these boundaries would support them in avoiding or assessing the risk of encountering hallucinations and misinformation. For teacher T12, contextualizing training data is essential for avoiding biased perspectives:

\begin{displayquote}
T12: \textit{If we're trying to look at a certain text with a certain like racial consideration in mind, then if I knew like ChatGPT \dots has never read books by a Black author then like that would probably change what I think it's capable of doing, you know, so I might not ask it that question in the first place.}
\end{displayquote}

More recently, biases may be attributed either to the underlying model or the added layers of restrictions upon ChatGPT output. Teachers pointed to the variety of filters in ChatGPT as an added source of uncertainty. One history teacher found that though content filters scrub openly hateful content, they may have also censored more controversial items in the Black Panther Party’s Ten-Point Program, leading to questions about the degree to which challenging topics in educational materials are sanitized. Several teachers described the experience of discovering that the interface no longer supported using ChatGPT to grade student work, prompting lowered confidence in its changing capabilities. 

Finally, nearly all teachers indicated limited knowledge of the ChatGPT interface and required prompting practices for effectively tailoring outputs to their needs. One teacher struggled to include a rubric as an input in the prompt while generating student writing samples according to rubric specifications. Teachers noted the lack of such practical information in available documentation and voiced preferences for resources to facilitate interactive exposure to ChatGPT.

\subsection{Capability Exploration and Experimentation}
To identify teachers' bespoke (learning task-specific) information needs, we engaged teachers in experimentation within the ChatGPT interface to explore its capabilities. In Table ~\ref{expgoals}, we present a taxonomy of teachers' experimentation behaviors. Their tests surfaced additional concerns about the process of prompt engineering and the pedagogical relevance of ChatGPT. Here we present observations about teachers' experimentation goals and prompting strategies, as well as their evaluation of ChatGPT outputs.

\subsubsection{Experimentation Goals}
Most teachers described their first experimentation goals as obtaining a general understanding of ChatGPT and the implications of its capabilities. They generated writing samples to elicit demonstrations of language understanding and capacity for personal authorship support. Teachers also recounted exploring the limitations of ChatGPT, searching for niche information, and provoking harmful responses.

In the education setting, teachers' early interactions with ChatGPT also often focused on verifying capabilities discussed in media sources: 
\begin{displayquote}
T7: \textit{Because I know teachers are curious about it \dots I've never done this, but they put things like “Make me a lesson plan on X”. Okay. So why don't I try that?}
\end{displayquote}
Teachers predominantly repeated documented use cases involving authorship, either of lesson plans or student essays, and simple information searching. Further, due to the common connection between ChatGPT and student cheating, teachers across all interviews reported an experimentation goal of assessing the risk of plagiarism. Teachers took the perspective of a student and entered prompts to generate personal narratives, argumentative essays, comprehension question responses, and citations. In doing so, teachers aimed to identify detectable characteristics of ChatGPT-generated language or evaluate outputs for flaws to share with students as a deterrent for using the tool for ineffective cheating. Across interviews, several teachers repeatedly gravitated back to this limited set of experimentation goals. 

When encouraged to think creatively about other applications for ChatGPT, many teachers explored prompts to optimize instructional tasks, preferring to avoid scenarios involving student interaction with the tool. Teachers experimented with compiling lists of resources, creating historical timelines and other reference documents, and generating summaries of textbook chapters. They used ChatGPT to generate lesson plans, writing samples, project ideas, comprehension questions, and parent emails:

\begin{displayquote}
T10: \textit{And then how can we, as teachers, can use it to make our jobs easier. Like if I need to write a sample response or an essay instead of me sitting down and spending an hour writing this sample response, can I have ChatGPT write me a draft and then like fix it to suit my needs?}
\end{displayquote}

In the second phase of interviews, teachers were guided to adopt the experimentation goal of augmenting student learning. By encouraging teachers to systematically consider the pedagogical motivations and student learning needs associated with potential use cases for ChatGPT, teachers experimented with automating feedback, simplifying the language of documents to student reading levels, and generating differentiated and standards-aligned assessment materials. Several teachers experimented with varying degrees of scaffolding for student writing. For example, teacher T17 shared an idea for transforming students' free writing, brainstorming, and class notes into organized outlines. 

Situating generative tasks in student learning also introduced specific requirements for outputs. Consequently, teachers were compelled to more carefully consider the construction of their inputs. Prompt engineering then became an additional experimentation goal as teachers navigated interactions to obtain more useful results.

\begin{table}[]
\resizebox{\columnwidth}{!}{%
\begin{tabular}{p{4cm} p{9cm} p{9cm}}
\toprule
\textbf{Experimentation Goal} & \textbf{Use of ChatGPT}                                   & \textbf{Examples}                                                                                                                                      \\
\midrule
Obtain General Understanding  & Demonstrations of Language Understanding                  & Generating simple sentences, haiku, style imitation                                                                                                   \\
                              & Personal Authorship Support                               & Generating cover letters, emails                                                                                                                       \\
                              & Exploring Limitations                                     & Searching for niche information, provoking harmful responses                                                                                           \\
Verify Described Capabilities & Instructional materials authorship support                & Generating lesson plans, essay prompts                                                                                         \\
                              & Student authorship support                                & Generating student essays, answers to comprehension questions                                                                                          \\
                              & Research simplification                                   & Information searching, text summarization                                                                                      \\
Assess Risk of Plagiarism     & Identifying characteristics of ChatGPT-generated language & Generating citations, generating personal narratives and other student essays, assessing capabilities to manipulate vocabulary, tone, and subjectivity \\
                              & Plagiarism detection                                      & Generate scores and reports identifying plagiarism                                                                                                     \\
Optimize Instructional Tasks  & Academic content authorship support                       & Learning objectives, lesson plans, project instructions, essay prompts, comprehension questions, writing samples                                       \\
                              & Administrative authorship support                         & Lesson plan documentation, parent emails, recommendation letters                                                                                       \\
                              & Research simplification                                   & Information searching, text summarization, resource compilation                                                                                        \\
                              & Analyzing student writing                                 & Grading, editing, proof-reading, generating feedback for student work                                                                                  \\
Augment Student Learning*      & Student authorship support                                & Generating ideas, essay outlines, paragraph completion from outline, diction                                                                           \\
                              & Analyzing student writing                                 & Proof-reading, generating feedback                                                                                                                     \\
                              & Research simplification                                   & Information searching, text summarization, resource compilation                                                                                        \\
                              & Content formatting                                        & Generating study materials, simplified explanations                                                                                                    \\
Prompt engineering*            & Instructional materials authorship support                & Changing the reading level of text, specifying or emphasizing output requirements \\
\bottomrule
\end{tabular}
}
\caption{Teachers' goals and behaviors observed in interviews for exploring the capabilities of ChatGPT. Goals marked with an asterisk were observed primarily in the second phase interviews involving guided experimentation.}
\label{expgoals}
\end{table}

\subsubsection{Prompting Strategies}
Across all interviews, teachers expressed reservations about their abilities to effectively interact with the open-ended ChatGPT interface. Teachers faced difficulties phrasing their input prompts. A few teachers worked around this discomfort by copying prompts from samples in online articles. Others initially overcame the cold start by approaching the tool from the perspective of a student to generate pieces of writing, but again struggled with prompting when tasked to consider other applications. A prominent source of frustration stems from teachers’ mental models connecting ChatGPT to other known tools. Several teachers compared the prompting process to that of using search engines:

\begin{displayquote}
T20: \textit{So I’m kind of stuck like I don't actually feel like I know really how to use the tool \dots it almost feels like I need to know how to engineer the right thing. I guess I'm comparing it a little bit to like a Google search like what's the best keywords to put in to get the right output.}
\end{displayquote}

Without researcher intervention, all teachers initially approached prompting with short inputs no longer than a single sentence. Though a few teachers described prior experiments in which more detailed prompts produced more usable outputs, these details often involved baseline contextual information such as grade level, subject area, and desired length of output text. 

In the revised interview protocol, teachers were encouraged to discuss specific student learning needs, and the prompting strategy of detailing contextual variables became more involved. When generating lesson materials, a few teachers specified student characteristics, describing student interests, feelings of belonging in school, and preferred modes of demonstrating understanding. Teachers incorporated pedagogical requirements, using terms from lesson materials. For example, teacher T19 prompted ChatGPT for feedback on student work specifically regarding identifying instances of indirect characterization and evaluating the degree to which students’ effective use of this writing skill builds suspense. Upon learning that ChatGPT retained the context of previous prompts and outputs, a few teachers began to approach prompting as an iterative process, introducing new information in extended follow-up prompts to refine outputs. For example, one teacher prompted ChatGPT first to generate a set of essay questions about a novel, then used subsequent prompts to revise questions to reflect the language of the Common Core state standards and appeal to specified student interests.

In the second phase of interviews, teachers considered the student experience of prompting, motivating them to rethink prior prompting practices. Though teachers had previously used the strategy of including age and grade level information to manipulate the vocabulary and sentence complexity of outputs, a few teachers expressed concern that the approach may not be appropriate for student-led interactions. When teacher T18 prompted ChatGPT to simplify the language of a reading passage for students of varying language proficiency, their initial approach produced a useful but unsettling response:

\begin{displayquote}
T18: \textit{I found this one, “Explain like I am 12” and \dots “Please simplify the language you use” \dots I feel like students could do that, but I also worry \dots these prompts basically concede a deficit view of their own knowledge.}
\end{displayquote}

In practice, teachers are tasked to design prompts that appropriately specify relevant contextual factors to support pedagogical objectives while accounting for student experiences. The difficulty of engineering such prompts is further compounded by limitations in the outputs of ChatGPT.

\subsubsection{Response Validation}
Many teachers encountered difficulties with the accuracy of information produced by ChatGPT. Experimentation with the use of ChatGPT for research surfaced gaps in training data, filters and deliberate omissions, and hallucinations. Teachers found factually incorrect summaries of linked articles and lesser known works of literature, and incomplete information about global political issues. A few teachers confronted fabricated citations while searching for resources and expressed frustration with ChatGPT’s inability to identify original sources of output content. 

Additionally, teachers noted the tendency for ChatGPT to produce generic responses. This behavior occurred inconsistently across similar tasks, exacerbating teachers’ uncertainties regarding prompting and expected output. Many teachers found the generated lesson plans to be basic and abstract. Several teachers observed that responses remained mainstream and conservative, limiting resource suggestions to works by well-known authors rather than those less frequently anthologized. In social studies classes, teachers noticed outputs containing hedging and qualifying language to avoid complex stances. 

Across all interviews, teachers noted the overwhelming length and complexity of outputs and found them inappropriate for student interactions. Several teachers noticed repetition and circular logic in generated feedback for student writing, noting that such language fails to provide process supports for students to implement suggestions. Teacher T9 expressed concern at the litany of criticisms generated in response to a prompt for feedback:

\begin{displayquote}
T9: \textit{This does not help developing writers. What helps them is getting positive feedback on what they're doing right and one or two key points to fix. This would shut somebody down.}
\end{displayquote}

Indeed, ChatGPT lacks pedagogical context. One teacher prompted ChatGPT to design a lesson plan with a constructivist lens but found the pedagogical construct present only in the language of the output rather than in the content or activities described. Several teachers prompted ChatGPT to generate writing prompts, essential questions, or comprehension questions but were disappointed in the outputs that could not distinguish the three categories of materials. In several interviews, ChatGPT did not recognize terms critical to instructional goals. While manipulating language to simplify instructions for English language learners, teacher T20 found that the outputs had changed the original intention behind the assignment: 

\begin{displayquote}
T20: \textit{When I’m trying to teach subject-specific vocabulary, I don’t want to take out words like “theme”. \dots Making something more simplistic isn’t necessarily like teaching what students should do. The first time I put it in there, it said “Give historical context about the poem” versus “Say a little bit about what the poem is about”. Those are entirely different things.}
\end{displayquote}

These demonstrated limitations of ChatGPT exacerbate barriers to meaningful adoption. They place further demands on teachers’ skillful prompting and ability to translate the educational domain to a technical interface. 

\subsection{Integration Support Needs}
Integrating ChatGPT in classroom practice requires a process of complex decision-making considering technical capabilities, instructional needs and objectives, and administrative procedures. However, teachers faced significant process barriers and persistent challenges with designing classroom processes surrounding ChatGPT and navigating its pedagogical role. 

\subsubsection{Evaluation Failures}
Teachers approached articles, blogs, and documentation without objectives for evaluating the information presented. Given disclosures of model limitations or evidence of low-quality outputs, teachers were unable to assess the severity of risks. Consequently, teachers displayed a highly forgiving attitude towards ChatGPT. 

Though outputs often contained factual errors or pedagogically inappropriate content, many teachers were willing to “let it slide”. Without expectations of the roles ChatGPT should play in the classroom and without a strategy for evaluating the quality of outputs, many teachers were quick to designate the tool as useful after affirming any level of utility in a single use case. Further, they were subsequently unmotivated to refine the output or find additional applications. Teacher T22 explained their contentment with personally revising undesired ChatGPT output:

\begin{displayquote}
T22: \textit{There's no rule to say like what ChatGPT spits out is what you have to use. So a lot of times you get mediocre responses or incorrect responses, or just kinda useless responses, but you can always iterate from them.}
\end{displayquote}

The tendency for teachers to compensate for unmet information needs by assuming personal responsibilities is also reflected in their responses to potential harms. Rather than seeking additional information or support, several teachers dismissed disclosures of biases and limitations in the model card as avoidable:

\begin{displayquote}
T12: \textit{Because that's stuff that we're familiar with \dots so as dangerous as it could be, it seems like the bell curve flattens really quickly. I don't know that I would ever ask it a question that would give it a chance to say something horrible, right?}
\end{displayquote}

The shifted accountability places an additional burden on teachers and creates a false sense of agency. As a result, teachers prematurely abandon their evaluation of model limitations. Model transparency, especially when set in a process derailed by unmet support needs, fails to meaningfully affect practice. 

\subsubsection{Classroom Procedural Adaptations}
Across interviews, most teachers framed their experiences with ChatGPT around “combating” its involvement in student cheating. For some, this involves added layers of monitoring, over-reliance on AI detectors, and re-introducing hand-written and timed in-class assessments, despite having previously discontinued their use. Other teachers lamented the pressure to adapt all assessments to avoid being what teacher T19 calls “cheatable”. Lacking formal guidance, teachers struggled with the design of new assessments. Teacher T16 expressed their frustration with abstract suggestions for teachers to adapt content:

\begin{displayquote}
T16: \textit{It's not useful, because it's like “Turn things project-based so they can't just plug it into ChatGPT”, or things like that that aren't necessarily like practical. There's not a lot of stuff available \dots that is like how to use it effectively.}
\end{displayquote}

Further, teachers struggled to define new classroom procedures to manage and monitor student use of ChatGPT. Though a few teachers shared their solutions, including asking students to print and annotate their chat records to justify their prompting choices, most voiced support needs. Several teachers noted that procedures permitting the use of ChatGPT often require additionally defining adapted evaluation criteria and academic standards. Across all interviews, teachers expressed their discomfort with navigating the added responsibility of significant procedural adaptations on an ad-hoc basis. 

\subsubsection{Pedagogical Trade-offs}
While some teachers lauded the tool as a “game-changer” in providing pre-service or early career teachers with sample lesson plans and administrative writing, others expressed concern about how responsibilities are divided. Teachers navigated the uncertain balance between finding creative applications for the unique technical affordances of ChatGPT and understanding the changing role of their own creativity. 
As teachers experimented with the ability of ChatGPT to optimize teachers’ tasks, they searched for moments of irreplaceable teacher involvement. Teacher T1 explained their hesitation with using ChatGPT to generate writing prompts:

\begin{displayquote}
T1: \textit{I enjoy doing that work, and I think it makes me a better teacher to have done that work and thought about like “how could this be like answered and addressed?”.}
\end{displayquote}

Teachers voiced a need for guidance regarding their role in the classroom upon integrating an open-ended interface with the technical capacity to perform numerous tasks that overlap significantly with those traditionally performed by teachers. The instructional role most jeopardized is that of assessing student understanding. Beyond fears of student cheating, student interactions with ChatGPT may reduce teachers’ opportunities to observe student learning processes. Further, applications of ChatGPT that aim to provide feedback and process supports personalize student learning within the interface and do the teaching work of responding to mistakes and misunderstandings. Such applications have the potential to both augment student learning and erode teacher agency for follow-up:

\begin{displayquote}
T11: \textit{As a teacher, as any type of teacher, not just English, you're trying to see the kids’ thinking. And I'm just worried that this actually creates a barrier to that, and I can't see if my kid like needs help or doesn't understand the concept, or even like if no one's getting it like. If everybody is kind of like making the same mistake, or pitfall, or whatever, then that's something for me to address \dots and affects what I'm gonna do with the next day even.}
\end{displayquote}

\section{Discussion} \label{sec: discussion}
Our interviews reveal important insights about teachers' experiences with the integration of ChatGPT. We observed their strategies for seeking and making sense of information, exploring technical capabilities, and evaluating applicability in their classroom contexts. Teachers face significant information gaps and support needs, limiting their abilities to interact with and find appropriate applications for ChatGPT. Here, we discuss the ways in which available static documentation inadequately addresses integration barriers and propose a framework for interactive model documentation that empowers teachers to navigate the interplay between pedagogical and technical knowledge. We raise open questions about accountability and implications for AI ethics and the broader education domain.

\subsection{Limitations of Model Documentation}
Designed to provide readers with an understanding of how models work to inform decision-making and enable the pursuit of workarounds for limitations~\cite{mitchell2019model}, model documentation seems to answer the call for AI literacy in education~\cite{schiff2022education}. However, while current AI documentation practices provide significant disclosures about model limitations, they have limited utility in the education domain. Their formatting and abstract content inadequately support teachers' effective integration of ChatGPT.

\subsubsection{Documentation is Static} That teachers found model documentation inaccessible and impractical is due largely to their narrow technical focus. However, the negative perceptions of documentation are influenced in part by the same frustrations that teachers expressed about guides and articles in educational publications: the text-based format. Static formatting is incompatible with teachers' preferred modes of consuming information about educational technologies, which include short videos of interface walk-throughs, hands-on personal exploration, classroom piloting with student feedback, and troubleshooting sessions with experts. Given well-known time constraints \cite{wepner2003three}, teachers shared their fatigue and disengagement about reading long texts with uncertain utility. 

\subsubsection{Documentation Restricts Active Discovery}
Though existing resources provide high-level advice and sample prompts that may inspire teachers \cite{openai_teachingguide}, these supports fail to scaffold teachers’ own creativity. In our interviews, teachers struggled to discover appropriate use cases for ChatGPT and repeatedly returned to the few use cases popularized by media narratives and documented samples, despite finding the resulting outputs generic and incompatible with their classroom practices. Teachers explored ChatGPT's capabilities through a limited set of generative tasks, primarily composing lesson plans and student essays. Overwhelmed by the open-ended interface, teachers faced a failure of imagination, following documented patterns or otherwise abandoning the integration process. In this way, they face what Ko et al. \cite{ko2004six} describe as design barriers, in which users struggle to identify the purposes to which a system can be applied, and selection barriers, in which users struggle to discover system behaviors. To conceptualize opportunities for integration, teachers require interaction with a broad range of generative abilities and opportunities to specify the use cases relevant to their pedagogical needs. 

Current documentation provides examples but fails to scaffold the necessary skills for teachers to modify applications to their unique contexts. We observed teachers' limited prompting strategies and challenges in navigating the ChatGPT interface. Though teachers often identify faults in the resulting outputs, they lack the technical knowledge of prompt engineering to achieve desired behaviors. Without guidance to effectively interact with the ChatGPT interface, teachers lose agency and perceive insurmountable challenges to integration.

\subsubsection{Documentation Lacks Domain-Relevance}
In the TPACK framework, effective integration of technology requires teachers to integrate technical knowledge with pedagogical and content knowledge \cite{mishra2019considering}. However, while resources document technical information and disclosures about ChatGPT, they critically lack support for teachers to contextualize the information in domain-relevant terms. Though teachers found disclosures about the limitations in dataset composition and the risk of misinformation and representational harms important, they could not imagine how such limitations would manifest in practice. In attempts to understand technical limitations in context, teachers asked questions about whether the training data contained specific textbooks or works of literature and about the comparative degree of risk certain content scenarios may incur for eliciting incomplete or inaccurate information. However, current documentation is too abstract to equip teachers to productively work around limitations. As a result, some teachers expressed overconfidence that they could easily identify and avoid disclosed limitations while others opted to avoid the use case entirely and strengthened their resolution against integrating ChatGPT. 

Further, by focusing on technical affordances and limitations, documentation neglects information gaps critical to technology integration in the education domain. Teachers require support navigating their adapted instructional roles and the design of new ChatGPT-integrated classroom procedures in the context of changing educational systems \cite{mishra2021contextualizing}. The use of ChatGPT in classrooms necessitates considerations for administrative oversight, classroom management practices, and assessment methods and standards. Yet, guidance for coordination among classroom roles and processes remains largely missing from documentation. Suggestions that teachers should simply review students' interactions with ChatGPT \cite{openai_faq} relegate teachers to the role of reacting to misuse rather than proactively defining practices.

\subsection{A Framework for Guided Experimentation}
Researchers have called for teachers to experiment and ``play'' with generative AI to develop new pedagogical techniques and discover system behaviors in the context of instructional goals \cite{mishra2019considering}. In support of this objective, documentation should equip teachers with the appropriate skills to engage in such experimentation. Resources should center the communication of technical information in the context of relevant learning scenarios, bridging technical and pedagogical knowledge \cite{celik2023towards}. Moreover, in order for documentation to orient and support teachers' use of ChatGPT, it must support interactivity \cite{crisan2022interactive}. To address these desired characteristics of documentation, we propose a layered framework for designing classroom applications of ChatGPT that augment student learning. The framework decomposes the open-ended interface of ChatGPT and defines five levels of considerations for developing an integration scenario: learning objectives, student needs, generative artifacts, prompting engineering, and classroom coordination. In ongoing research, we apply the framework with teachers in a series of design interviews and make initial recommendations for its integration in the design of a tool allowing teachers to explore---rather than read about---the technical capabilities of ChatGPT. 

\begin{figure}[h]
\includegraphics[width=\textwidth]{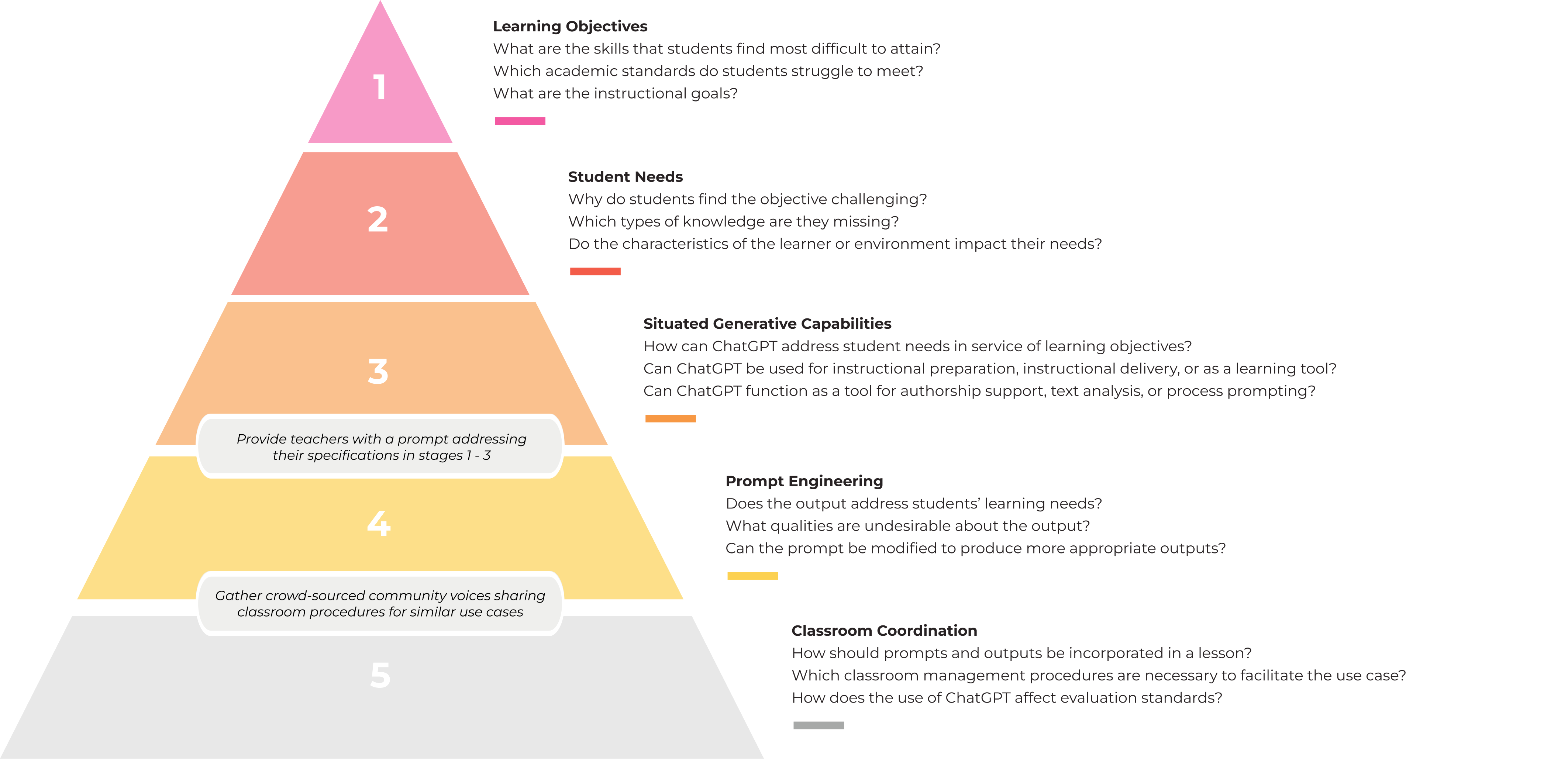}
\caption{Stages of considerations to guide teachers' exploration of classroom applications of ChatGPT. Each stage introduces multiple routes to inspire experimentation.}
\label{fig:framework}
\end{figure}

\subsubsection{Learning Objectives}
Because current documentation focuses on the technical affordances of ChatGPT, we observed teachers struggling to fit documented capabilities into their classroom practice. Consequently, their approach to integration relied on a few, often redundant, use cases poorly aligned to their instructional goals. 
To address the narrow scope of applications, the first dimension of the framework encourages teachers to ground their exploration in the context of learning objectives, without considering ChatGPT. We ask teachers to identify the problems of their domain and specify the knowledge and skills, defined by academic standards, that students find most difficult to attain. For example, in writing instruction, teachers may point to students' abilities to support claims with logical reasoning and relevant evidence or to establish and maintain a formal style \cite{commoncore_writing}. 

\subsubsection{Student Needs}
Though our interview participants represented a diversity of school environments, including elite private schools, large underfunded public schools with significant English language learner populations, and remedial institutions, the characteristics of their unique classroom contexts were rarely reflected in their approach to integration. Instead, teachers noted the importance of manually revising the generic outputs of ChatGPT based on their knowledge of their students' abilities and interests. To better reflect these factors, the second dimension of the framework asks teachers to specify the cognitive and contextual needs of students for each learning objective identified in the previous stage, again without considering ChatGPT. In addition to identifying relevant features of the classroom setting, we ask teachers to consider the reasons for which students may find the objective challenging and the types of knowledge \cite{de1996types} that students may lack. For example, students may struggle to support claims with relevant evidence due to language barriers in reading comprehension, difficulties identifying the main message of passages, lack of interest or engagement in the topic, or insufficient exemplars.

\subsubsection{Situated Generative Capabilities}
Discovering applications of ChatGPT requires creativity, but we observed teachers exploring applications of ChatGPT through a limited set of generative tasks. To help teachers imagine additional possibilities, the next dimension invites teachers to consider a broad range of generative artifacts to explore each problem space defined through the previous stage of the framework. We define these options through the instructional roles of technology and the generative roles of ChatGPT. The use of technology in schools can be expressed through three broad categories: technology for instructional preparation, technology for instructional delivery, and technology as a learning tool \cite{inan2010factors}. For each category, ChatGPT may function as a tool for authorship support, text analysis, and process prompting. For example, to address student difficulties identifying the main message of passages, ChatGPT may be used in instructional preparation by generating passages and comprehension questions as formative assessments, highlighting the sentences within input passages that best represent the main messages, or suggesting how the teacher might explain and correct a student's misinterpretation of the main message of a passage. To address the same challenge, ChatGPT may be used as a learning tool by students to generate a sentence expressing the main idea of an input passage, generate explanations detailing the contribution of each sentence to the overall message of the passage, or create a conversational tutor asking a series of comprehension questions leading to students identifying the main message of the passage.

\subsubsection{Prompt Engineering}
Due to the significant information gaps and misconceptions about prompting observed in our interviews, in this stage, our recommended design for the interactive tool involves supplying teachers with an initial prompt based on their specifications in the previous three stages. Beginning with this initial prompt, we scaffold teachers' understanding of prompt engineering by first asking teachers to identify undesired qualities in the model output. For example, the format of the output may be inappropriate in style and length, or the content may inadequately address student needs. We propose suggesting prompt modifications on-demand to address teacher-identified output problems. In doing so, we break down the complexity of prompt engineering by framing it as a process of iteratively improving outputs. For example, if the output contains vocabulary above the reading level of students for a student-facing use case, teachers are informed to add the specification of language level to their prompt. While many solutions involve additional specifications in prompting, other problems present an opportunity to convey technical knowledge. For example, teachers exploring the use of ChatGPT as a conversational partner may need to be informed of interface features such as the ability to leverage chat log history to orchestrate a sequence of prompts, and teachers identifying false or biased information in outputs may need to be reminded of model limitations. 

\subsubsection{Classroom Coordination}
The orchestration of roles between teachers and AI systems is critical \cite{roschelle2020ai}, but complex and unsolved. In our interviews, teachers repeatedly expressed concerns about the impact of ChatGPT integration on classroom management and the standards by which students are assessed. Like the design of use cases and prompts, the design of classroom procedures involving ChatGPT is influenced by context. However, while scholars express optimism for such procedures to evolve from teachers' increasing familiarity with the tool \cite{mishra2023chatgpt}, experimentation alone does not sufficiently address the urgent problem of coordinating instructional roles, administrative oversight, classroom management, and lesson procedures in the learning environment. Our recommended design for the final stage of the framework involves crowd-sourced accounts detailing teacher experiences with ChatGPT-integrated classroom procedures, surfaced selectively to teachers based on alignment with specifications in the first three stages of the framework.  

In summary, the presented framework is pedagogy-driven, student-focused, and interactive. By enabling teachers' discovery of possible applications of ChatGPT in contextually relevant learning scenarios and scaffolding teachers' abilities to tailor interactions to their specifications, the framework addresses key support needs revealed through our interviews. Note that the framework does not address all barriers to integration and is intended to complement---not replace---static documentation and ethical disclosures. However, it provides teachers with a guided approach to the open-ended interface of ChatGPT and an opportunity to infer through experimentation the way such tools function.

\subsection{Open Questions}
Though we identify teachers' information gaps and support needs about ChatGPT, our findings reveal deeper unknowns about the complexity of its integration into education. To provide teachers with adequate training and support, the roles and responsibilities of various stakeholders require definition, calling into question the degree to which AI developers can be held accountable. We additionally question the effect of model documentation and disclosures on the ethical deployment of AI systems, and we discuss implications for curricular reform, access, and equity.

\subsubsection{Accountability}
Currently, the burden of adapting to and effectively utilizing ChatGPT falls almost unilaterally on teachers. They are expected to update their instructional competencies and develop new classroom procedures by integrating new technical knowledge in pedagogical skills \cite{mishra2023chatgpt}. These urgent efforts required to address the disruption of ChatGPT compound teachers' demanding workloads \cite{butt2005secondary, beck2017weight}, contributing to well-documented patterns of burnout and attrition \cite{brill2008stopping}. 

One way to reduce teachers' burden is to distribute responsibilities to administrators, professional development programs, and teacher training programs. These domain stakeholders have traditionally assumed the responsibility of connecting technical and pedagogical knowledge by training teachers in TPACK competencies for technology integration \cite{brinkley2018learning}. Indeed, recent research has discussed the implications for training programs and districts' efforts to better prepare teachers for integrating ChatGPT \cite{whalen2023chatgpt}. 

However, the responsibilities of OpenAI remain undefined. ChatGPT lacks a critical feedback mechanism commonly found in other educational technology integrations. As teachers described in our interviews, the adoption processes for other educational technology tools involve representatives from companies who are responsible for receiving and responding to bug reports, contributing to the content of professional development, and hosting troubleshooting sessions. Teachers' discontent about broken functionality or features that fail to address classroom needs is eventually reflected in district purchasing decisions. As a result, conventional educational technology companies have a stake in listening to teachers' feedback and making suitable adjustments to products. 

In contrast, the creators of ChatGPT are not accountable for responding to the problems raised by educators and parents. Though the U.S. Government's AI Bill of Rights calls for users to "access a person who can quickly consider and remedy problems [they] encounter" \cite{aibillofrights}, such aspirations are not legally binding. There is no mechanism for flagging low-quality and problematic content, demanding transparency about student data, or requesting administrative controls essential to the education domain. Instead, OpenAI states in a recent FAQ \cite{openai_faq} that use by minors is against the terms of use, but without changes to the interface to regulate use, such statements are inconsistent with practitioner realities.
 
\subsubsection{Implications for Ethical AI}

Despite industry efforts to require model documentation and ethical statements \cite{ashurst2022ai} to address fairness, accountability, and transparency of AI systems, current standards for their contents fail to ensure ethical deployments. Disclosures about dataset composition, performance metrics, and representational harms narrowly construe fairness as a small set of harms that motivate algorithmic fixes \cite{boyarskaya2020overcoming}. However, these abstract documents are insufficient to address the range of impacts from generative AI systems \cite{solaiman2023evaluating} and critically lack domain-specific considerations. They neglect the full scope of unpredictable context-dependent harms affecting end-users \cite{boyarskaya2020overcoming, veale2018fairness}. We observed these harms in teachers' significant information gaps and support needs in response to the deployment of ChatGPT in education. Teachers face practical challenges managing student use, ensuring student safety, navigating curricular disruptions, and urgently seeking new technical literacies. As ChatGPT obscures teachers' abilities to observe and respond to students' understanding, and fears of cheating result in an over-reliance on AI detectors with serious academic consequences, the harms of ChatGPT directly impact student learning. 

Further, documentation and ethical disclosures are posthumous, articulating harms as an afterthought to deployment. Such documents deflect and downplay negative societal impacts, shifting blame to improper inputs or malicious use, declaring limitations to caution end-users, and delegating the undefined task of mitigating limitations to practitioners while effectively absolving companies and AI developers of responsibility \cite{liu2022examining}. In response, researchers have proposed an ethics and societal impact review process as a prerequisite to research funding \cite{bernstein2021ethics}. Had this process been implemented in the case of ChatGPT, it may have involved a limited and controlled release through risk-assessed sectors, modified interfaces appropriate to partner domains, and early collaborations with domain stakeholders to anticipate usability traps and support needs. However, in the case of ChatGPT, a teaching guide of sample prompts and FAQ answers \cite{openai_teachingguide} is released in response to months of practitioner outcry. 

\subsubsection{Implications for Education Systems}
The popular narrative comparing ChatGPT to calculators \cite{calculator} was reflected in our interviews. The debate about calculators in math instruction placed curricular content and the design of educational standards into question. Advocates called for replacing arithmetic-driven lessons that emphasize rote skills and reimagining the content that is relevant to teaching. 
Though teachers in our interviews expressed similar perspectives about ChatGPT, they were often unable to articulate the qualities of curricular reform. They emphasized the need to eliminate "cheatable" assignments but were waiting for another authority to design the assessments that would be implemented in their place. These unknowns arise in part from a flawed metaphor---ChatGPT is unlike a calculator in that the capabilities of generative AI are far greater than carrying out mechanical operations---and in part from ongoing technical advancements that continue to change the skills students may need in their futures. 

One change in response to ChatGPT involves new requirements for teaching technology and addressing the presence of AI in the lives of students. In our interviews, teachers discussed their challenges with finding ways to introduce ChatGPT to students and discuss its limitations. Districts facing similar unknowns have responded by designing AI curriculum \cite{edweek_ai_curriculum} and expanding computer science education \cite{edweek_ai}. However, the curricular implications of preparing students for a future in which every domain of human activity is affected by generative AI \cite{li2020artificial} remains unknown.

Finally, open questions remain regarding access and equity. Teachers in our interviews often referenced equal access to opportunity as the reason for adopting ChatGPT, motivated by exposing their students to tools and resources available to their peers in other educational contexts. However, our study participants also acknowledged that they were more engaged in learning about ChatGPT and other new technologies than their peers. The consequences of varying teacher abilities to limit the disruption of ChatGPT and teach students productive uses of the tool are yet to be seen.

\section{Conclusion}
Effective integration of educational technology relies on teachers’ abilities to take pedagogical advantage of technical affordances. Yet, in the case of ChatGPT, teachers are inadequately supported in developing these competencies. By observing teachers' information-seeking and exploration of ChatGPT, we identify significant information gaps. While model documentation informs educators about ChatGPT's capabilities and discloses limitations in model behaviors, it lacks domain relevance, and teachers are unable to operationalize the information in their classroom practices. Our findings raise important questions about the ethical deployment of AI systems and the necessary supports to address consequent harms.

\bibliographystyle{ACM-Reference-Format}
\bibliography{99_refs, 07_techadoption, 08_aiedu, 09_softwaredocs, 10_aidocs}

\end{document}